\documentclass[a4paper,10pt,oneside,preprintnumbers,showpacs,superscriptaddress,nofootinbib,amsmath,amssymb,floats,floatfix,showkeys, nofootinbib,notitlepage]{revtex4-2}
\usepackage{orcidlink}
\usepackage{comment}
\usepackage{lipsum}
\usepackage{graphicx}
\usepackage{subfigure}
\usepackage{palatino}
\usepackage{sans}
\usepackage{hyperref}
\hypersetup{colorlinks=true,linkcolor=blue,urlcolor=blue,citecolor=blue}
\usepackage[toc,page]{appendix}
\usepackage[normalem]{ulem}
\usepackage{adjustbox}
\usepackage{latexsym}
\usepackage{amsmath}
\usepackage{amssymb}
\usepackage{amsfonts}
\usepackage{dcolumn}
\usepackage{bm}
\usepackage{tikz}
\usepackage{bigints}
\usepackage{array,tabularx,multirow,booktabs}
\usepackage[tracking=true]{microtype}
\SetTracking{}{500}
\SetTracking{encoding={*}, shape=sc}{40}
\UseRawInputEncoding 
\allowdisplaybreaks

\begin{document} \sloppy

\title{Imprints of a gravitational wave through the weak field deflection of photons}

\author{Reggie C. Pantig
\orcidlink{0000-0002-3101-8591}}
\email{rcpantig@mapua.edu.ph}
\affiliation{Physics Department, Map\'ua University, 658 Muralla St., Intramuros, Manila 1002, Philippines}

\author{Ali \"Ovg\"un
\orcidlink{0000-0002-9889-342X}
}
\email{ali.ovgun@emu.edu.tr}
\affiliation{Physics Department, Eastern Mediterranean University, Famagusta, 99628 North Cyprus via Mersin 10, Turkiye.}

\begin{abstract}
This paper investigates the novel phenomenon of gravitational lensing experienced by gravitational waves traveling past a Schwarzschild black hole perturbed by a specific, first-order, polar gravitational wave. We utilize the Gauss-Bonnet theorem, uncovering a topological contribution to the deflection of light rays passing near the black hole. We demonstrate that the deflection angle can be determined by analyzing a region entirely outside the light ray's path, leading to a calculation based solely on the parameters of the perturbing wave (Legendre polynomial order, $l$, and frequency, $\sigma$). This approach offers a unique perspective on gravitational lensing and expands our understanding of black hole interactions with gravitational waves.
\end{abstract}


\maketitle


\section{Introduction} \label{intr}
A black hole is an enigmatic region of spacetime where nothing, not even light, can escape due to the immense warping of spacetime caused by an exceptionally massive object, such as a star. Black holes remain some of the most intriguing and mysterious phenomena in the universe, presenting significant challenges to our comprehension of the laws of physics. On the other hand, Albert Einstein's theory of general relativity (GR) predicted the existence of gravitational waves, which are ripples in the fabric of spacetime resulting from compact masses that accelerate. The LIGO Scientific Collaboration confirmed their existence in 2015 \cite{LIGOScientific:2016aoc}. These waves are produced when two massive objects, such as black holes or neutron stars, orbit each other and eventually merge. As they revolve around one another, they emit energy in the form of gravitational waves that travel outward in all directions at the speed of light. The direct detection of gravitational waves in the landmark event GW150914 \cite{LIGOScientific:2016aoc} by the Virgo Scientific and LIGO collaborations not only validated Einstein's GR but also heralded a new era of discovery in the fields of astrophysics, cosmology, and astronomy. Since then, several other gravitational wave events have been detected, including binary black hole mergers and binary neutron star mergers \cite{LIGOScientific:2017vwq,KAGRA:2022osp}. By comparing gravitational wave signals with theoretical models, we can glean valuable information about astrophysical objects, such as their spins, masses, and other parameters. Furthermore, detecting gravitational waves promises to enhance our understanding of black holes and validate various theories of gravity \cite{Ghosh:2022xhn}.

Gravitational lensing is an excellent tool in gravitational physics, enabling us to study the properties of massive objects in the universe. The phenomenon occurs when light from a distant source is bent as it passes through a gravitational field, such as a black hole or a galaxy. This bending causes the image of the source to appear distorted or magnified, creating multiple images of the same object. The effect of gravitational lensing arises from the warping of spacetime by mass and energy, in accordance with Einstein's theory of general relativity. Although the effect is usually small, it can be significant when the gravitational field is very strong \cite{Virbhadra:1999nm,Virbhadra:2002ju,Adler:2022qtb,Bozza:2001xd,He:2020eah,Junior:2021svb,Cunha:2019hzj,Cunha:2016bjh}.  In the realm of astrophysics, the ability to measure the distances of celestial objects is critical for gaining insights into their fundamental characteristics. Yet, Virbhadra showed that it is possible to estimate an upper limit on the compactness of massive, unobservable objects simply by analyzing the relativistic images without prior knowledge of their masses and distances \cite{Virbhadra:2022ybp}. Moreover, Virbhadra identified a parameter of distortion that leads to summing all the singular gravitational lensing images to disappear with a sign (Schwarzschild lensing has been used to test this in both weak and strong gravitational fields regimes, \cite{Virbhadra:2022iiy}). On the other hand, the Gauss-Bonnet theorem (GBT) is a powerful tool in gravitational physics as it allows us to relate the curvature of spacetime to its topology, which is the study of the properties of a space that are preserved under continuous transformations, such as stretching or twisting. The theorem states that for any smooth, compact, orientable two-dimensional surface embedded in a three-dimensional space, the total curvature equals $2 \pi$ times the Euler characteristic of the surface. In gravitational lensing, this theorem is used by Gibbons and Werner to calculate the deflection angle of light as it passes through a gravitational field, such as that of asymptotically flat black holes \cite{Gibbons:2008rj}. By considering the topological properties of the surface of the black hole, we can obtain an expression for the deflection angle that depends on the mass and angular momentum of the black hole, in addition to the impact parameter of the incoming light. This approach relied on GBT and the optical geometry of metric describing the black hole, where the source and receiver of light rays are situated in asymptotic regions. This method was later extended to axisymmetric spacetimes by Werner \cite{Werner2012}, who utilized Nazim’s osculating manifold construction. Following this, Ishihara and colleagues developed this technique for finite-distances, specifically for cases with large impact parameters \cite{Ishihara:2016vdc} and by Ono et al \cite{Ono:2017pie}, who explicitly considered the contribution of the one-form rotation to the geodesic curvature and accounted for finite distance corrections.  Li et al. extended the generalized optical metric method to the generalized Jacobi metric method for massive particles \cite{Li:2019vhp,Li:2020dln,Li:2020wvn}. Huang and Cao generalized the Gibbons-Werner method for weak deflection angle \cite{Huang:2022soh}. This method has been used extensively in the literature to study the effects of black holes on the deflection of light. It has been shown to be a powerful tool for understanding the properties of these mysterious objects \cite{Ovgun:2018fnk,Ovgun:2019wej,Ovgun:2018oxk,Javed:2019ynm,Ono:2017pie,Li:2020dln,Li:2020wvn,Belhaj:2022vte, Javed:2023iih, Javed:2023IJGMMP, Javed:2022fsn, Javed:2022gtz,Qiao,Huang:2023iog,Huang:2022soh,Atamurotov:2022knb,Guo:2022hjp,Zhang:2021ygh,Fu:2021fxn,Fu:2021akc,Lessa:2020imi,Ovgun:2023ego,Pulice:2023dqw,Lobos:2022jsz}.

Studying gravitational wave effects on gravitational lensing is an interesting avenue to explore, given their potential role in studying black holes and verifying various alternative theories of gravity. However, it is proven difficult to obtain a gravitational wave solution due to the complexity of perturbing the Einstein field equation. Nevertheless, the paper in \cite{1981Xanthopoulos} obtained a specific gravitational wave solution that satisfies Einstein's equations expanded up to the first order in $\epsilon$.

The solution provides a means to investigate the unique gravitational waves' effect on the dynamics of a test particle in black hole spacetime. Such a particular gravitational wave revealed the causes the motion of a timelike test particle to become non-integrable, leading to the emergence of chaotic phenomena \cite{Letelier:1996he}, which differs from those observed in the absence of gravitational waves. This chaotic behavior is expected to manifest in the photons' motion, resulting in interesting effects on black hole lensing. Hence, this paper aims to explore the impact of this gravitational wave on the deflection angle of black holes in the weak field regime.

We organize the paper as follows: In Sect. \ref{sec2}, we briefly introduce the Schwarzschild black metric perturbed by gravitational waves. We explored the weak deflection angle via GBT in Sect. \ref{sec3}. Finally, in Sect. \ref{sec4}, we conclude the paper. Through-out the paper, we set $G = c = 1$, implying that the mass is geometrized. The metric signature used in this study is $(-,+,+,+)$.

\section{Schwarzschild black hole perturbed by gravitational wave} \label{sec2}
The Schwarzschild metric is the simplest black hole metric that can be obtained from the vacuum solution of the Einstein field equation. If a particular gravitation wave perturbs this black hole, it was shown that the modification of the metric is given as \cite{1981RSPSA.378...73X,Wang:2019skw}
\begin{equation} \label{metric}
    ds^2 = (g_{\mu\nu} + \epsilon h_{\mu\nu})dx^\mu dx^\nu.
\end{equation}
\begin{equation} \label{metric2}
    ds^2 = -A(t,r,\theta)dt^2 + B(t,r,\theta)dr^2 + C(t,r,\theta)d\theta^2 + D(t,r,\theta)d\phi^2,
\end{equation}
where
\begin{align} \label{metcoef}
    A(t,r,\theta) &= f(r)\left(1 + \epsilon x P_l(\cos\theta) \cos(\sigma t) \right), \nonumber \\
    B(t,r,\theta) &= f(r)^{-1}\left(1 + \epsilon y P_l(\cos\theta) \cos(\sigma t) \right), \nonumber \\
    C(t,r,\theta) &= r^2\left[1 + \epsilon \left(z P_l(\cos\theta) + w \frac{d^2}{d\theta^2}P_l(\cos\theta) \right) \cos(\sigma t) \right], \nonumber \\
    D(t,r,\theta) &= r^2 \sin^2(\theta)\left[1 + \epsilon \left(z P_l(\cos\theta) + w \frac{d}{d\theta}P_l(\cos\theta) \cot(\theta) \right) \cos(\sigma t) \right].
\end{align}
In Eq. \eqref{metcoef}, $P_l$ is defined as the Legendre polynomial. As for the functions $w, x, y,$ and $z$, one finds solutions from $ R_{cdb}^c + \epsilon R_{cdb}^c h^{ab} = 0 $ as follows
\begin{align}
    f(r) &= 1 - \frac{2M}{r}, \quad x = pq, \quad y = 3Mq, \quad z = q(r - 3M), \nonumber \\
    w &= rq, \quad p = M - \frac{M^2+\sigma^2 r^4}{r - 2M}, \quad q = \frac{\sqrt{f(r)}}{r^2}.
\end{align}

We should note that this special type of gravitational wave is not the one found by LIGO \cite{LIGOScientific:2016aoc,LIGOScientific:2017vwq,KAGRA:2022osp}, which is short-lived. Theoretically, an isolated, non-rotating black hole is imagined to be passed through by a gravitational wave described by Eq. \ref{metcoef}, which is assumed to last longer in terms of time period. The most notable analysis is found in \cite{Wang:2019skw}, where they studied in detail how the black hole shadow will behave under this perturbation. Here, we extend the analysis by examining the effect of the gravitational wave on the deflection angle in the weak field limit in a simplified manner by considering time slices and restriction along $\theta_o = \pi/2$.

\section{Weak deflection angle} \label{sec3} 
In this section, we shall investigate the parameter $\epsilon$ in conjunction with the weak deflection angle denoted by $\hat{\alpha}$. To this aim, we exploit the GBT \cite{Gibbons:2008rj} expressed as
\begin{equation} \label{eGBT}
    \iint_T KdS+\sum\limits_{i=1}^N \int_{\partial T_{i}} \kappa_{\text{g}} d\ell+ \sum\limits_{i=1}^N \Theta_{i} = 2\pi\eta(T).
\end{equation}
Here, $K$ is defined as the Gaussian curvature, $dS$ is the measured areal surface, $\Theta_i$, and $\kappa_g$ are the jump angles and geodesic curvature of $\partial T$, respectively. We denote $d\ell$ as the arc length measure. Its utilization concerning null geodesics along the equatorial plane implies that the Euler characteristic should be $\eta(T) = 1$. Should the integral be evaluated through the unbounded surface area delimited by the light ray, it was shown in Ref. \cite{Ishihara:2016vdc} that Eq. \eqref{eGBT} reduces to
\begin{equation} \label{eIshi}
    \hat{\alpha}=\phi_{\text{RS}}+\Psi_{\text{R}}-\Psi_{\text{S}} = -\iint_{_{\text{R}}^{\infty }\square _{\text{S}}^{\infty}}KdS.
\end{equation}
Here, $\phi_{RS} = \phi_\text{R} - \phi_\text{S}$ is the equatorial separation angle between the light source S and receiver R (the observer), $\phi_\text{S}$ and $\phi_\text{R}$ are the respective positional angles, and $_{R}^{\infty }\square _{S}^{\infty}$ is the defined integration domain. The utilization of the above formula in this investigation is precluded since Eq. \eqref{metcoef} is indeed non-asymptotically flat, which is attributed to the presence of the gravitational wave parameter. Nonetheless, it has been demonstrated in \cite{Li:2020wvn} that by changing the integration domain of the light rays to the photonsphere instead of the path extending to infinity, Eq. \eqref{eIshi} can be reformulated into a form applicable to spacetimes that are not asymptotically flat (like the black hole spacetime with cosmological constant $\Lambda$):
\begin{equation} \label{eLi}
    \hat{\alpha} = \iint_{_{r_\text{ps}}^{R }\square _{r_\text{ps}}^{S}}KdS + \phi_{\text{RS}}.
\end{equation}
For us to determine the expressions for $K$ and $dS$, consider the metric in Eq.\eqref{metric2}. Here, our focus will exclusively be the light deflection along the equatorial plane. Notably, $P_l$ is contingent upon $\theta$, yet when $\theta = \pi / 2$, solely the even orders in $l$ for $P_l$ yield a dimensionless quantity. Given our interest lies in the deflection of photons, the optical metric can readily be derived when $ds^2 = 0$. That is,
\begin{equation}
    dl^2=g_{ij}dx^{i}dx^{j}
    = \left(\frac{B(t,r)}{A(t,r)}dr^2+\frac{D(t,r)}{A(t,r)}d\phi^2\right).
\end{equation}

In addition, the time dependence of the metric Eq. \eqref{metric2} implies that $E/\mu$ ($\mu = 1$ for time-like geodesics) can no longer become a constant of motion. Nevertheless, $L/\mu$ remains a constant of motion, owing to its independence on the $\phi$ coordinate. Despite these conditions, the Hamiltonian formalism does not constrain the expression of the impact parameter to be written as
\begin{equation}
    b = \frac{L}{E} = \frac{D(t,r)}{A(t,r)}.
\end{equation}
Therefore, given a slice of $t$, a particular value for $b$ is determined given $r$. Additionally, the orbit equation can be readily expressed as
\begin{equation} \label{eorb}
    \frac{dr}{d\phi } = \left[ {\frac {D(t,r)  }{B(t,r)  } \left( {\frac {D(t,r) }{A(t,r)  {b}^{2}}}-1 \right) }\right]^{1/2}.
\end{equation}
With the substitution $u = 1/r$, we obtain
\begin{align} \label{eorb2}
    F(u) \equiv \left(\frac{du}{d\phi}\right)^2 
    &= \frac{1}{b^2}-u^2+2mu^3 \nonumber \\
    - {\frac {\epsilon\,\cos \left( \sigma\,t \right) P_l \left( {u}^{4}{b}^{2}-{\sigma}^{2}-2\,{u}^{2} \right) }{u{b}^{2}}} 
    &+ {\frac {\epsilon\,mP\cos \left( \sigma\,t \right)  \left( 9\,{u}^{4}{b}^{2}+{\sigma}^{2}-12\,{u}^{2} \right) }{{b}^{2}}}.
\end{align}

We use $u = 1/r$, commonly employed in Newtonian celestial mechanics. Subsequently, employing an iterative approach, the objective is to ascertain $u$ as a function of $\phi$, which is determined as
\begin{equation} \label{euphi}
    u(\phi) = \frac{\sin(\phi)}{b}+\frac{(1+\cos^2(\phi))m}{b^2}+ {\frac { \left( {b}^{2}{\sigma}^{2}+1 \right) P_l \epsilon \cos \left( \sigma\,t\right) }{2\,{b}^{2}}} + \frac{P_l \epsilon m \left(b^{2} \sigma^{2}-1\right) \cos \! \left(\sigma  t \right)}{b^{3}}.
\end{equation}

\begin{figure*}
    \centering
    \includegraphics[width=0.48\textwidth]{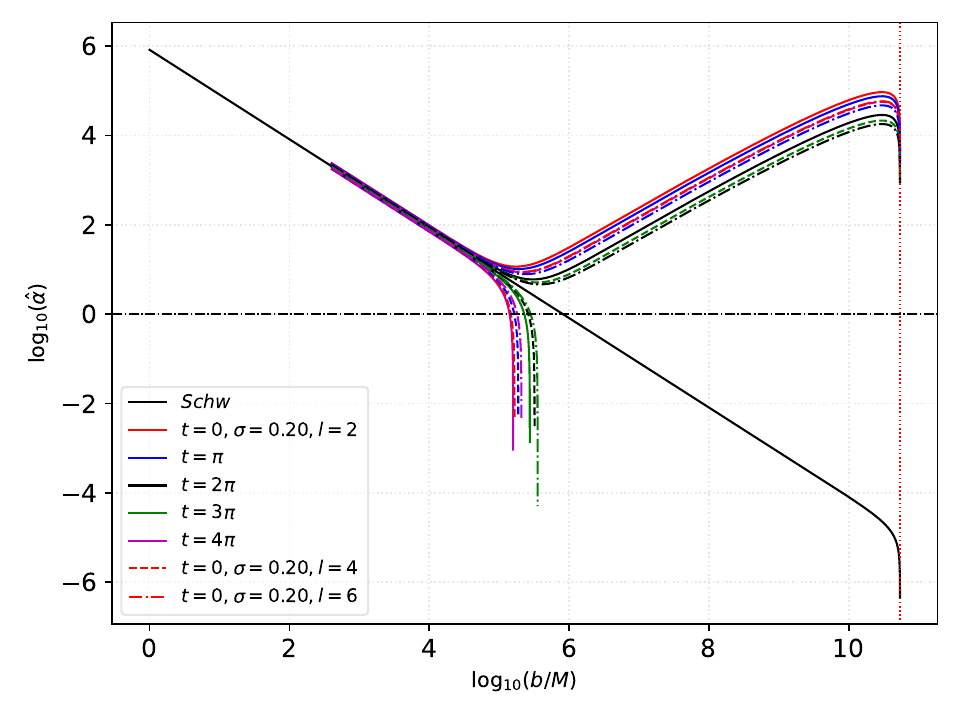}
    \includegraphics[width=0.48\textwidth]{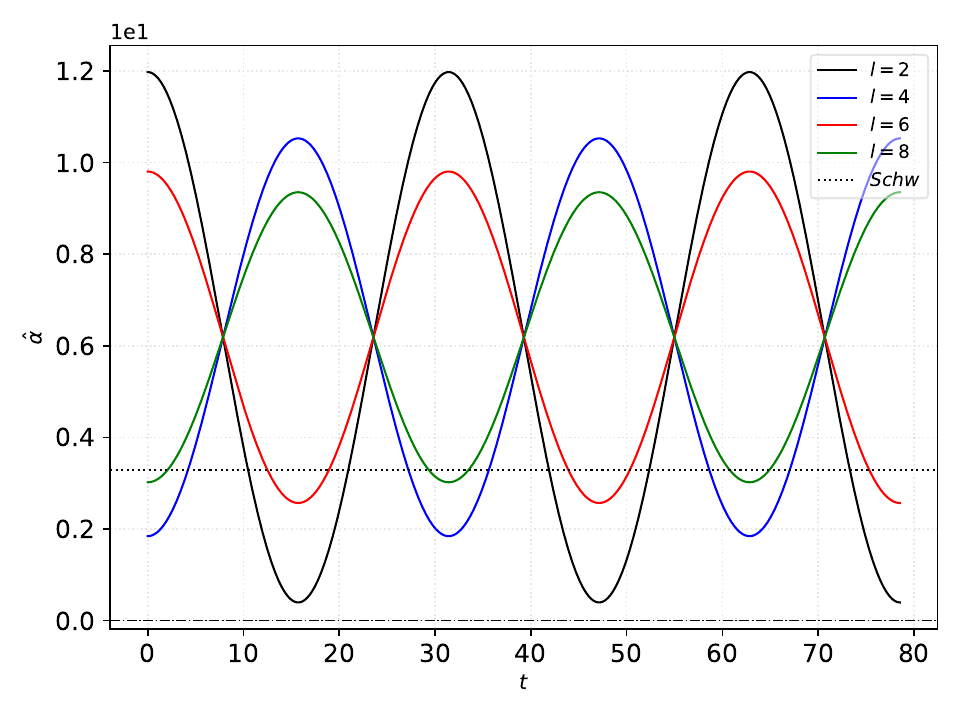}
    \caption{Left: Weak deflection angle in a log-log plot as $b/m$ varies. The vertical red dotted line represents Earth's distance from M87*, which is $16.8$ Mpc. The horizontal black dash-dot line is $\hat{\alpha} = 10^0$. Right: Weak deflection angle of photons at $b/m = 10^{5.125}$ as $t$ varies. In both figures, we assumed that $u_\text{R} = u_\text{S} = m / 16.8 \text{ Mpc}$, $\sigma = 0.20$, and $\epsilon = 10^{-9.5}$.
    }
    \label{wda_smbh}
\end{figure*}

We can derive the Gaussian curvature through the expression
\begin{align}
    K=-\frac{1}{\sqrt{g}}\left[\frac{\partial}{\partial r}\left(\frac{\sqrt{g}}{g_{rr}}\Gamma_{r\phi}^{\phi}\right)\right],
\end{align}
since $\Gamma_{rr}^{\phi} = 0$. Next, the optical metric's determinant is given as \cite{Gibbons:2015qja,Li:2019vhp}
\begin{equation}
    g=\frac{B(r)C(r)}{A(r)^2}(E^2-\mu^2 A(r))^2.
\end{equation}
If there is an analytical solution to the radius of the photon's unstable circular orbit $r_\text{co}$, it can be gleaned that \cite{Li:2020wvn}
\begin{equation}
    \left[\int K\sqrt{g}dr\right]\bigg|_{r=r_\text{co}} = 0,
\end{equation}
which results to
\begin{equation} \label{gct}
    \int_{r_\text{co}}^{r(\phi)} K\sqrt{g}dr = -\frac{A(r)\left(E^{2}-A(r)\right)C'-E^{2}C(r)A(r)'}{2A(r)\left(E^{2}-A(r)\right)\sqrt{B(r)C(r)}}\bigg|_{r = r(\phi)},
\end{equation}
where $A'$ and $C'$ are differentiation with respect to the radial coordinate $r$. Then, $\hat{\alpha}$ can be obtained as \cite{Li:2020wvn},
\begin{align} \label{eqwda}
    \hat{\alpha} = \int^{\phi_\text{R}}_{\phi_\text{S}} \left[-\frac{A(r)\left(E^{2}-A(r)\right)C'-E^{2}C(r)A(r)'}{2A(r)\left(E^{2}-A(r)\right)\sqrt{B(r)C(r)}}\bigg|_{r = r(\phi)}\right] d\phi + \phi_\text{RS}.
\end{align}
With the help of Eq. \eqref{euphi} in Eq. \eqref{gct}, we find that
\begin{align} \label{gct2}
    \left[\int K\sqrt{g}dr\right]\bigg|_{r=r_\phi} &=-\phi_\text{RS} -\frac{2m (\cos (\phi_\text{R})-\cos (\phi_\text{S})}{b} \nonumber\\
    &-{\frac {\ln  \left( \csc \left( \phi_\text{R} \right) - \csc \left( \phi_\text{S} \right) -(\cot \left( \phi_\text{R}\right) - \cot \left( \phi_\text{S}\right)  \right) \cos \left( \sigma\,t \right) P_l b{\sigma}^{2}\epsilon}{2}} \nonumber \\
    &+{\frac {\cos \left( \sigma\,t \right)  \left( \phi_{RS} \left( 2\,{\sigma}^{2}{b}^{2}+1 \right) +(\cot \left( \phi_\text{R} \right)-\cot \left( \phi_\text{S} \right))  \left( 1-2\,{\sigma}^{2}{b}^{2}- \left( \cos \left( \phi_\text{R} \right)-\cos \left( \phi_\text{S} \right)  \right) ^{2} \right) \right) P_l\epsilon\,m}{2\,{b}^{2}}}.
\end{align}
Next, the solution for $\phi$ can be obtained. The results are
\begin{align} \label{ephi}
    \phi_\text{S} &=\arcsin \! \left(b u_\text{S} \right)+\frac{\left(b^{2} u_\text{S}^{2}-2\right) m}{b \sqrt{-b^{2} u_\text{S}^{2}+1}}-\frac{\epsilon  \cos \! \left(\sigma  t \right) P_l \left(\sigma^{2} b^{2}+1\right)}{2 b \sqrt{-b^{2} u_\text{S}^{2}+1}} \nonumber\\
    &-\frac{\left(\sigma^{2} b^{2}-1\right) \cos \! \left(\sigma  t \right) P_l \epsilon  m}{\sqrt{-b^{2} u_\text{S}^{2}+1}\, b^{2}}, \nonumber\\
    \phi_\text{R}&=\pi-\arcsin \! \left(b u_\text{R} \right)-\frac{\left(b^{2} u_\text{R}^{2}-2\right) m}{b \sqrt{-b^{2} u_\text{R}^{2}+1}}+\frac{\epsilon  \cos \! \left(\sigma  t \right) P_l \left(\sigma^{2} b^{2}+1\right)}{2 b \sqrt{-b^{2} u_\text{R}^{2}+1}} \nonumber\\
    &+\frac{\left(\sigma^{2} b^{2}-1\right) \cos \! \left(\sigma  t \right) P_l \epsilon  m}{\sqrt{-b^{2} u_\text{R}^{2}+1}\, b^{2}}.
\end{align}
Noting the basic trigonometric relations $\cos(\pi-\phi_\text{S})=-\cos(\phi_\text{S})$ and $\cot(\pi-\phi_\text{S})=-\cot(\phi_\text{S})$, we find the following expressions:
\begin{align} \label{ecos}
    \cos(\phi) &= \sqrt{1-b^{2}u^{2}}-\frac{u \left(b^{2} u^{2}-2\right) m}{\sqrt{-b^{2} u^{2}+1}} + \frac{u \cos \! \left(\sigma  t \right) \left(b^{2} \sigma^{2}-5\right) \epsilon}{4 \sqrt{-b^{2} u^{2}+1}} \nonumber \\
    &+ \frac{\epsilon  m P \cos \! \left(\sigma  t \right) \left(-2+b \left(3 b \,u^{2}-2 u \right)+b^{2} \left(-2+b \left(3 b \,u^{2}+2 u \right)\right) \sigma^{2}\right)}{2 \left(-b^{2} u^{2}+1\right)^{\frac{3}{2}} b^{2}}
\end{align}
$\cot(\phi_\text{S})$ as
\begin{align} \label{ecot}
    \cot(\phi) &= \frac{\sqrt{1-b^{2}u^{2}}}{bu}-\frac{\left(b^{2} u^{2}-2\right) m}{b^{3} u^{2} \sqrt{-b^{2} u^{2}+1}} + \frac{\epsilon  \left(\sigma^{2} b^{2}+1\right) P \cos \! \left(\sigma  t \right)}{2 b^{3} \sqrt{-b^{2} u^{2}+1}\, u^{2}} \nonumber \\
    &+ \frac{\epsilon  m P \cos \! \left(\sigma  t \right) \left(-2+b \left(3 b \,u^{2}+u \right)+b^{2} \left(-2+b \left(3 b \,u^{2}-u \right)\right) \sigma^{2}\right)}{\sqrt{-b^{2} u^{2}+1}\, b^{5} u^{3} \left(b^{2} u^{2}-1\right)}
\end{align}
and $\ln(\csc(\phi_\text{S})-\cot(\phi_\text{S}))$ as
\begin{align} \label{eln}
    &\ln(\csc(\phi)-\cot(\phi)) = \ln\left(\frac{1}{bu}\right)+\ln(1-\sqrt{1-b^2u^2}) + \frac{m \left(b^{2} u^{2}-2\right)}{\sqrt{-b^{2} u^{2}+1}\, b^{2} u} \nonumber\\
    &-\frac{m P \epsilon  \cos \! \left(\sigma  t \right) \left(-2+b \left(3 b \,u^{2}+2 u \right)+b^{2} \left(-2+b \left(3 b \,u^{2}-2 u \right)\right) \sigma^{2}\right) \left(b^{2} u^{2}+\sqrt{-b^{2} u^{2}+1}-1\right)}{2 \left(\sqrt{-b^{2} u^{2}+1}-1\right) b^{4} u^{2} \left(b^{2} u^{2}-1\right)^{2}}.
\end{align}
Finally, with the help of Eqs. \eqref{ecot} through \eqref{eln} in combination with Eq. \eqref{gct2}, we obtained the approximated $\hat{\alpha}$ of light ray weak field deflection with added consideration to the finite distance effects of the source and the receiver:
\begin{align} \label{ewda}
    \hat{\alpha} &\sim \frac{2m}{b}\left(\sqrt{1-b^{2}u_\text{R}^{2}}+\sqrt{1-b^{2}u_\text{S}^{2}}\right)
    - \cos \! \left(\sigma  t \right) P b \,\sigma^{2} \epsilon \left[\mathrm{arctanh}\! \left(\sqrt{-b^{2} u_\text{R}^{2}+1}\right) + \mathrm{arctanh}\! \left(\sqrt{-b^{2} u_\text{S}^{2}+1}\right) \right] \nonumber \\
    & - \frac{2 m P \epsilon  \cos \! \left(\sigma  t \right)}{b^{3}} \Bigg\{ \frac{1}{4} \left[ \left(-3+2 b^{2} \left(2 u_\text{S}^{2}+\sigma^{2}+2 u_\text{R}^{2}\right)\right) \left(\sqrt{-b^{2} u_\text{R}^{2}+1}+\sqrt{-b^{2} u_\text{S}^{2}+1}\right)\right] \nonumber \\
    &+ \frac{\sigma^{2} b^{2}}{4}+\frac{1}{8} \left[ \arcsin \! \left(b u_\text{R} \right) b u_\text{R} + \arcsin \! \left(b u_\text{S} \right) b u_\text{S} \right] \Bigg\}.
\end{align}
The result above is interesting and confirms that Eq. \eqref{metric2} is indeed non-asymptotically flat due to the existence of $\mathrm{arctanh}$ terms in the second term. It only means that both $u_\text{R}$ and $u_\text{S}$ cannot equal zero. Nevertheless, if both are very close to zero, we can still reformulate Eq.\eqref{ewda} into its far approximation form:
\begin{align} \label{ewda2}
    \hat{\alpha} &\sim \frac{4m}{b} -\cos \! \left(\sigma  t \right) P b \,\sigma^{2} \epsilon \left[\mathrm{arctanh}\! \left(\sqrt{-b^{2} u_\text{R}^{2}+1}\right) + \mathrm{arctanh}\! \left(\sqrt{-b^{2} u_\text{S}^{2}+1}\right) \right] \nonumber \\
    &- \left[ \frac{m P \left(2 \sigma^{2} b^{2}-3\right) \cos \! \left(\sigma  t \right) \epsilon}{b^{3}}\right] \left( \frac{1}{u_\text{R}} + \frac{1}{u_\text{S}}\right)
\end{align}
Lastly, we plot the exact expression Eq. \eqref{ewda} as shown in Fig. \ref{wda_smbh}, where we also included the Schwarzschild case for comparison. In the left panel, we observe some interesting results as we consider different time slices for different orders in $l$. We retain the gravitational wave frequency to $\sigma = 0.20$. Looking around impact parameter $b/m \sim 3.0$, we can see that the value of $\hat{\alpha}$ fluctuates very close below and above the Schwarzschild case. The variation increases as $b/m$ gets larger up to $\sim 5.1$. The effect of the fluctuation disappears at very large $b/m$ as $\hat{\alpha}$ is shown is increasingly deviate from the Schwarzschild case dramatically. Such an effect is only seen for $t=0$ up to $t=2\pi$. We set $b/m = 10^{5.125}$ in the right panel to demonstrate the fluctuations as time progresses.

\section{Conclusion} \label{sec4}
This paper investigated a Schwarzschild black hole's weak gravitational lensing phenomenon, influenced by a gravitation wave due to a distinctive polar gravitational perturbation derived from Einstein's equations. Our main method exploits the Gauss-Bonnet theorem, where the integration domain spans from the photonsphere up to the source and receiver location, to admit spacetime metrics that are not asymptotically flat.

Our results indicated that certain $\hat{\alpha}$ fluctuations arise for light having $b/m$ around the range of estimated values $\sim 2.4$ to $\sim 5.2$. Interestingly, since the weak deflection angle is connected to lensed images, such as the existence of Einstein rings, one might possibly observe some pulsation of the lensed image. The deviation to the Schwarzschild case caused by the pulsation of the lensed image might hint at an alternative possibility of detecting gravitational waves using astronomical observation aside from Earth-based detection.

Research prospects include investigating other gravitational wave model and their possible effect and detection through the weak or strong field regime.

\section{ACKNOWLEDGMENTS}
A. {\"O}. and R. P. acknowledge the networking support of the COST Action CA18108 - COST Action CA21106 - COSMIC WISPers in the Dark Universe: Theory, astrophysics and experiments (CosmicWISPers), Quantum gravity phenomenology in the multi-messenger approach (QG-MM), the COST Action CA21136 - Addressing observational tensions in cosmology with systematics and fundamental physics (CosmoVerse), and the COST Action CA22113 - Fundamental challenges in theoretical physics (THEORY-CHALLENGES). We also thank TUBITAK and SCOAP3 for their unwavering support of this project.

\bibliography{ref}
\bibliographystyle{apsrev}

\end{document}